\documentstyle[aps,multicol,graphicx]{revtex}

\begin{document}

\title{The excess wing in the dielectric loss of glass-forming ethanol: A
relaxation process}
\author{R. Brand, P. Lunkenheimer, U. Schneider, and A. Loidl}
\address{Experimentalphysik V, Universit\"{a}t Augsburg, D-86135 Augsburg, Germany}
\date{29.02.2000; submitted to Phys. Rev. B}
\maketitle

\begin{abstract}
A detailed dielectric investigation of liquid, supercooled liquid, and
glassy ethanol reveals a third relaxation process, in addition to the two
processes already known. The relaxation time of the newly detected process
exhibits strong deviations from thermally activated behavior. Most
important, this process is the cause of the apparent excess wing, which was
claimed to be present in the dielectric loss spectra of glass-forming
ethanol. In addition, marked deviations of the spectra of ethanol from the
scaling proposed by Dixon and Nagel have been detected.
\end{abstract}

\pacs{PACS numbers: 77.22.Gm, 64.70.Pf}

\section{INTRODUCTION}

Dielectric spectroscopy plays an important role in the investigation of the molecular dynamics in glass-forming
materials. Due to the exceptionally wide frequency/time window accessible with this method, broadband dielectric
spectra reveal the large variety of processes governing the dynamic response above and below the glass temperature
$T_{g}$ \cite{Conte}. Among these, the microscopic origin of the process leading to the so-called excess wing (also
called ''high-frequency wing'' or ''Nagel-wing'') is still unclear. In the frequency-dependent dielectric loss
$\varepsilon ^{\prime \prime }(\nu )$, the excess wing shows up as an additional contribution to the high-frequency
power law of the $\alpha $-relaxation peak ($\varepsilon ^{\prime \prime }\sim \nu ^{-\beta }$). It can be reasonably
well described by a second power law, $\varepsilon ^{\prime \prime }\sim \nu ^{-b}$, with $b<\beta $
\cite{wingpow,Menon}. The excess wing, which was already noted in the early work of Davidson and Cole \cite{CD}, was
found in a variety of glass-forming materials \cite{wingpow,Menon,Di90a,Ku99}. In another class of glass-forming
materials, at frequencies above the $\alpha $-peak frequency $\nu _{p}$, a shoulder or even a second peak shows up,
giving clear evidence for a second faster relaxation process \cite{Ku99,Johari}, usually termed $\beta $-process
\cite{rem}. By considering the detailed molecular structure of a material, $\beta $-processes sometimes can be
ascribed to intramolecular degrees of freedom, especially in polymeric systems. But a systematic investigation of
various low molecular-weight glass-formers where such contributions can be excluded, revealed that these so-called
Johari-Goldstein $\beta $-relaxations may be inherent to glass-forming materials in general \cite{Johari}.
Consequently more fundamental reasons for their occurrence have been proposed \cite{Johari,Ngaibeta}.

Commonly it is assumed that the excess wing and $\beta $-relaxations are different phenomena \cite{Di90a,Ku99} and the
existence of two classes of glass-formers was proposed - ''type A'' without a $\beta $-process but showing an excess
wing and ''type B'' with a $\beta $-process \cite{Ku99}. However, very recently, by performing dielectric aging
experiments below $T_{g}$, we found strong hints that in glass-forming glycerol and propylene carbonate a second
relaxation process (called ''$\beta $-relaxation'' in the following, but see the remarks below) is the origin of the
excess wing observed in these materials \cite{wingbet}. Due to the fact that the relaxation time of this process is
relatively close to the $\alpha $-relaxation time \cite{Lunkihab}, only the high-frequency flank of the corresponding
relaxation peak becomes visible, thereby appearing as excess wing. In addition, for the orientationally disordered
phase of cyclo-octanol, a $\beta $-relaxation was unmasked as the origin of an apparent excess wing \cite{LP} by
simply extending the frequency range of dielectric spectra to higher frequencies \cite{oct}. Consequently, in
\cite{wingbet} it was proposed that $\beta $-relaxations may provide an explanation for the excess-wing phenomenon in
general and that the difference between ''type A'' and ''type B'' systems may simply be a different temperature
evolution of the $\beta $-process.

In an effort to check this notion, we initiated a systematic investigation of glass-forming materials that have been
reported to exhibit well-pronounced excess wings. In the present work we report results on glass-forming ethanol.
There is a large variety of publications concerning the disordered phases of ethanol in recent literature
\cite{eth1,Mi98,Be98,JR99}. Aside of the well-known common-life applications of this substance, the recent scientific
interest in ethanol was mainly triggered by the fact that it can be prepared both in a structurally disordered and
plastic-crystalline phase \cite{ethplas,eguchi}. In plastic crystals the centers of mass of the molecules form a
crystalline lattice but the molecules are orientationally disordered. Studies in both disordered phases of ethanol
have contributed to our understanding of the importance of orientational degrees of freedom in the supercooled state
of matter \cite {eth1,Mi98,Be98,JR99}. A variety of dielectric studies of this material have appeared\
\cite{Mi98,Be98,JR99,Hassion,Garg,Barthel,Kindt,Stickdis,Stick2}. In \cite{Be98,JR99} for the first time dielectric
loss spectra in ethanol were shown that revealed a significant excess contribution at the high-frequency flank of the
$\alpha $-peak. It was claimed as another example of the excess wing known from other glass-formers \cite{Be98}. In
the present work we report results of a detailed dielectric investigation of liquid, supercooled liquid and glassy
ethanol in a frequency range $3{\rm \mu Hz}<\nu <500{\rm MHz}$ and at temperatures $40{\rm K}<T<230{\rm K}$. Compared
to the earlier publications \cite{Mi98,Be98,JR99} we provide additional data at lower frequencies and at temperatures
which are difficult to access due to an enhanced crystallization tendency. Most important, we have obtained more
precise results in the excess-wing region at high frequencies, $\nu >1{\rm MHz}$. This allows for the unequivocal
detection of a third relaxation process in ethanol, which is responsible for the excess-wing feature observed in
earlier works.

\section{EXPERIMENTAL DETAILS}

High-precision measurements of the dielectric permittivity in the frequency range $100{\rm \mu Hz}\leq \nu \leq 1{\rm
MHz}$ were performed using a Novocontrol alpha-analyzer. Results at lower frequencies, down to $3\times 10^{-6}{\rm
Hz}$, were collected with a time domain technique. For the frequency range above $1{\rm MHz}$ a Hewlett-Packard HP4291
impedance analyzer was employed. For details the reader is referred to \cite{exp}. For cooling,\ the sample capacitor
was inserted into a closed cycle refrigerator or a ${\rm N}_{2}$ gas-heating system. The temperatures were precisely
measured by a Si-diode, completely inserted into one of the capacitor plates. Ethanol with a purity of $\geq 99.9\%$
was used for the measurements. In the region of $100-130{\rm K}$ ethanol exhibits an enhanced crystallization
tendency. Spectra at $T\leq 110{\rm K}$ were obtained after passing this region with rapid cooling rates ($8{\rm
K/min)}$ and subsequent heating to the desired temperature. In addition, during the cooling run, spectra at $114{\rm
K}$ and $118{\rm K}$ were collected with a reduced number of frequencies per decade and shorter integration time. In
this way the temperature drift during these frequency sweeps could be reduced to less than $0.5{\rm K}$.

\section{RESULTS AND DISCUSSION}

\subsection{The spectra}

Figure 1 shows the dielectric loss spectra for the structurally disordered phases of ethanol. The dominating $\alpha
$-relaxation peak shifts through the frequency window with temperature. The overall behavior is in good agreement with
the findings in \cite{Mi98,Be98,JR99}. However, at low temperatures, $T\leq 100{\rm K}$, there is a discrepancy to the
peak positions reported in \cite{Mi98,JR99} of up to one decade while a good match with the results in \cite{Be98} can
be stated. At $T=110{\rm K}$ the $\alpha $-peak exhibits a shoulder at $\nu <\nu _{p}$ and its amplitude is reduced.
This finding can be ascribed to a partial transition

\begin{figure}
\begin{center}
\includegraphics[angle=-90,clip,width=16cm]{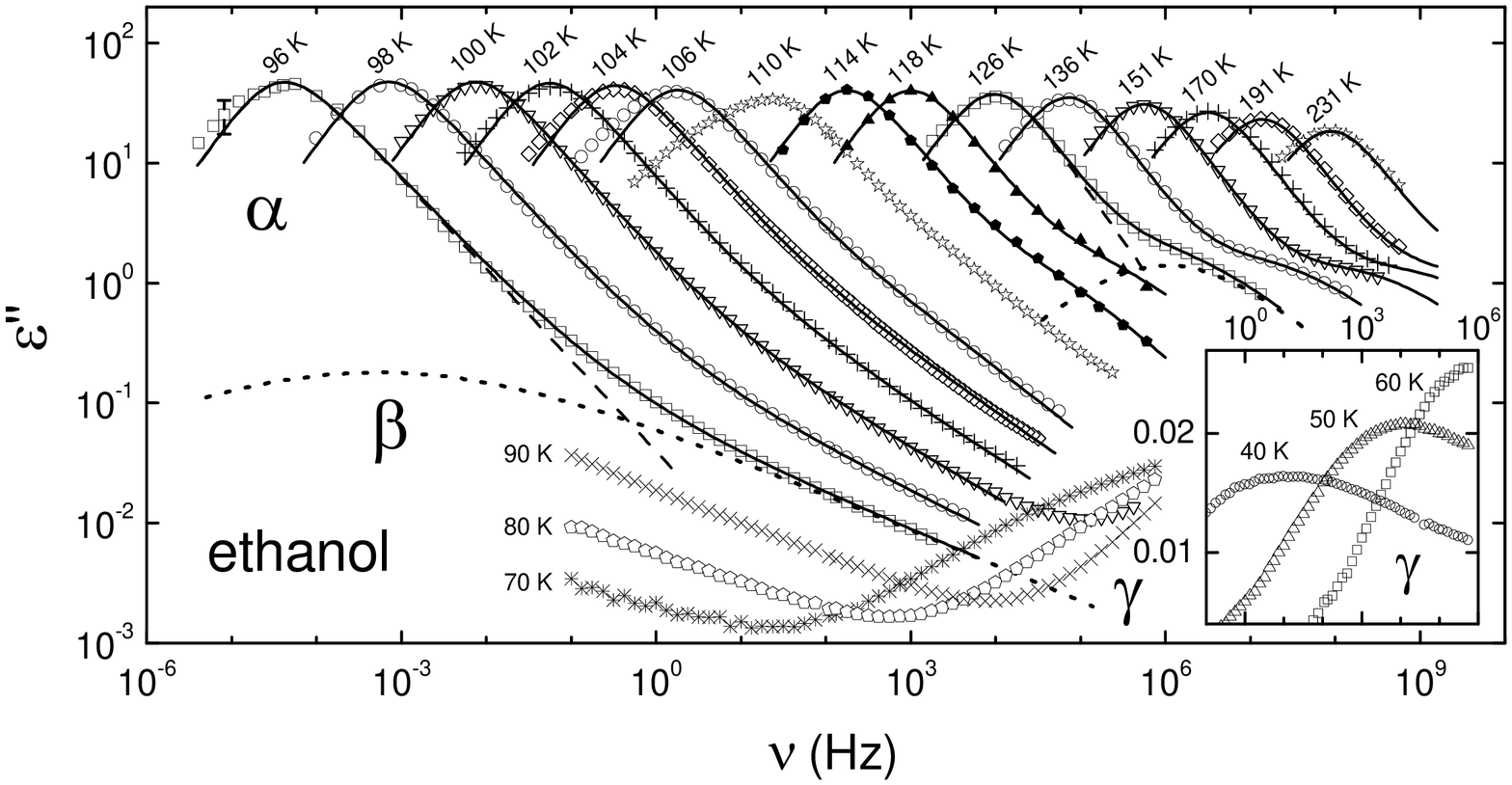}
\end{center}
\caption{Frequency-dependent dielectric loss of glass-forming ethanol for various temperatures. The solid lines are
fits with the sum of a CD and a CC function, performed simultaneously for $\protect\varepsilon ^{\prime }(\protect\nu
)$ and $\protect\varepsilon ^{\prime \prime }(\protect\nu )$. For $86{\rm K}$ and $126{\rm K}$, the dashed lines show
the two constituents of the fits. The inset gives a separate view of the low-temperature results.} \label{fig1}
\end{figure}

\begin{figure}[tbp]
\begin{center}
\includegraphics[clip,width=8cm]{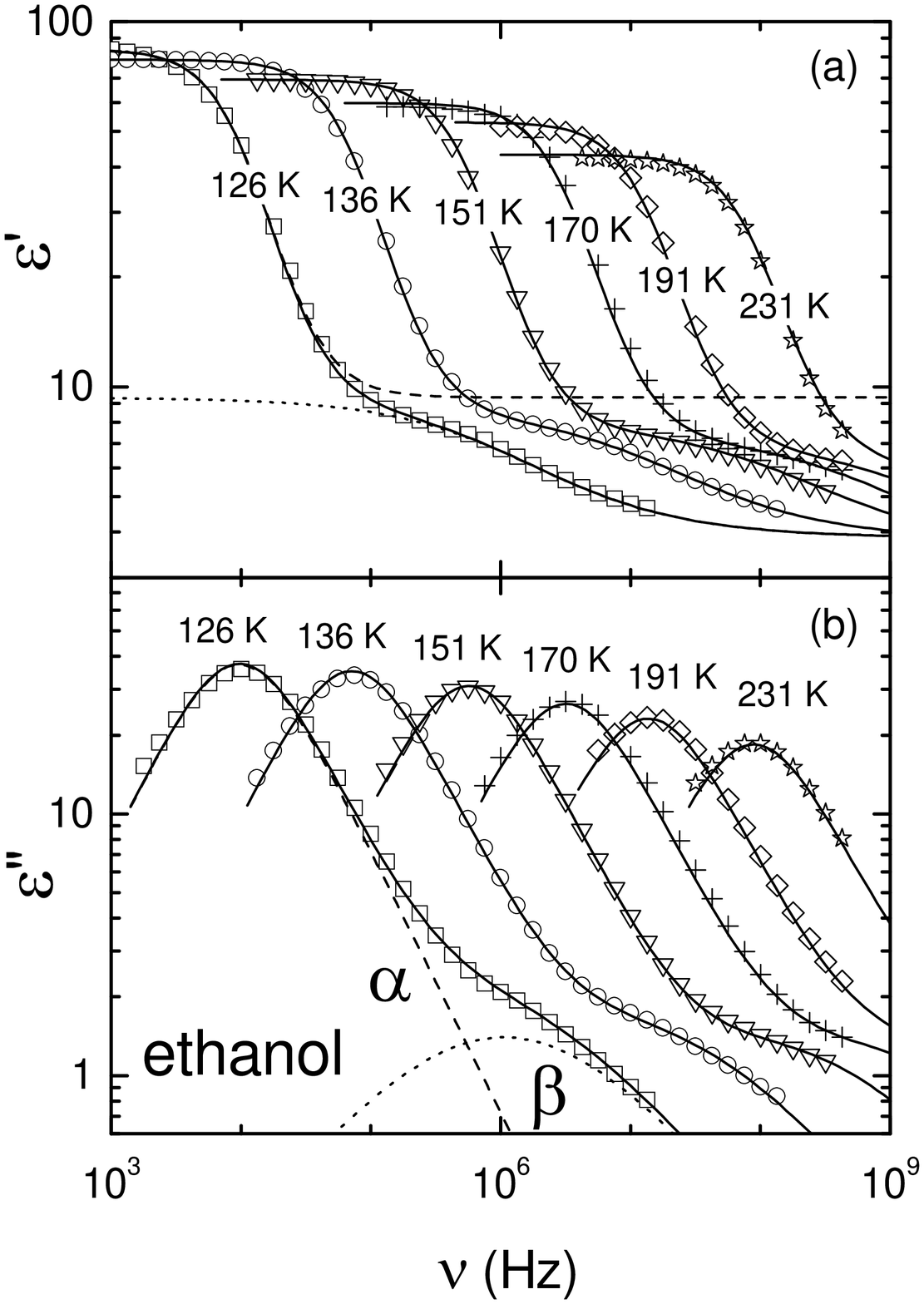}
\end{center}
\caption{Dielectric constant (a) and loss (b) spectra of liquid and supercooled ethanol at frequencies above $1{\rm
kHz}$. The solid lines are fits with the sum of a CD and a CC function, performed simultaneously for
$\protect\varepsilon ^{\prime }(\protect\nu )$ and $\protect\varepsilon^{\prime \prime }(\protect\nu )$. For $126{\rm
K}$ both components of the fit are shown separately (CD: dashed line, CC: dotted line).} \label{fig2}
\end{figure}

\noindent of the sample into the plastic crystalline phase which occurred during the approach of this temperature from
below, similar to the observations made in \cite{Be98} at $105{\rm K}$. In good agreement with the findings in
\cite{Be98,JR99}, at $T\leq 110{\rm K}$ an excess contribution to the $\nu ^{-\beta }$-power law of the $\alpha $-peak
is observed. This feature was interpreted as excess wing \cite{Be98}. At temperatures below $T_{g}=97{\rm K}$ and
frequencies above this ''excess wing'', $\varepsilon ^{\prime \prime }(\nu )$ starts to rise again. Finally, below
$60{\rm K}$, a second relaxation-peak shifts into the frequency window (inset of Fig. 1) in agreement with earlier
reports \cite{Be98,JR99}. The upturn of $\varepsilon ^{\prime \prime }(\nu )$ observed at $100{\rm K}$ and $\nu
>100{\rm kHz}$ indicates that this relaxation process is present at $T>T_{g}$, too.

The most important result of the present work is revealed at temperatures $T\geq 126{\rm K}$: At frequencies about two
decades above $\nu _{p}$ a shoulder shows up, i.e. $\varepsilon ^{\prime \prime }(\nu )$ exhibits a downward curvature
[see also Fig. 2(b)]. This finding clearly indicates the presence of a {\it third} relaxation process in ethanol. This
notion is confirmed by the observation of well-developed additional relaxation steps in $\varepsilon ^{\prime }(\nu )$
as demonstrated in Fig. 2(a). The loss-spectra at temperatures $T\geq 126{\rm K}${\rm , }presented in \cite{Be98}
obviously do not provide sufficient precision or sufficiently high frequencies to allow for a detection of this
relaxation. Instead only a feature resembling an excess wing was observed. In \cite{JR99} the data at $\nu
>1{\rm MHz}$ are reported for $T\geq 160{\rm K}$\ and in a semilogarithmic plot only, which prevented the observation
of this relaxation. In the following the newly detected relaxation will be termed ''$\beta $-relaxation'' and the
relaxation observed in the glass state\ (denoted as ''$\beta $-relaxation'' in \cite{Be98,JR99}) will be called
''$\gamma $-relaxation''. This nomenclature is simply intended to take account of the succession of these relaxations
in the frequency window, without making a statement about their physical origin.

Figure 1 strongly suggests that the $\beta $-relaxation, resolved as a shoulder at high temperatures, develops into
an\ apparent excess wing at low temperatures. Here the situation is similar to that suggested by us for the
explanation of the excess wing in glycerol and propylene carbonate \cite{wingbet}: Due to the close vicinity of
$\alpha $- and $\beta $-relaxation times, only the high-frequency flank of the $\beta $-peak shows up as an apparent
excess wing. The solid lines in Figs. 1 and 2 are fits with the sum of the empirical Cole-Davidson (CD) and Cole-Cole
(CC) functions \cite{boettcher}, very similar to the approach in \cite{JR99} where a sum of two CD functions was used
for fits at lower temperatures \cite{remsum}. Instead of the second CD function, in the present work a CC function was
chosen, which is known to often provide a satisfactory parameterization

\begin{figure}[tbp]
\begin{center}
\includegraphics[clip,width=10cm]{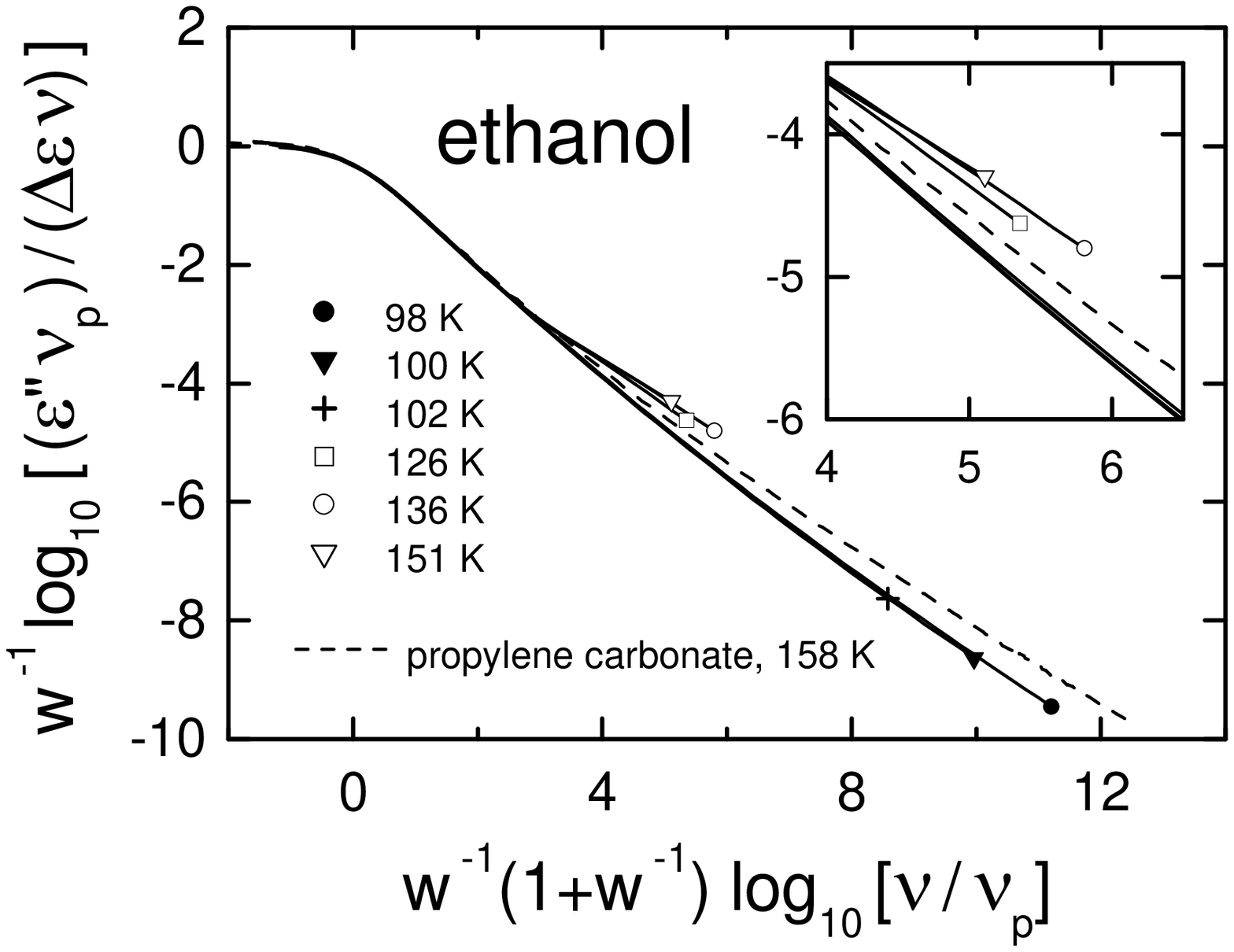}
\end{center}
\caption{Scaling plot of the dielectric loss in glass-forming ethanol after Dixon and Nagel \protect\cite{Di90a} at
selected temperatures (solid lines). The symbols indicate the highest-frequency point for the different temperatures.
The $\protect\varepsilon ^{\prime \prime }(\protect\nu )$-data of the increase towards the $\protect\gamma
$-relaxation were not used for the scaling. The dashed line shows the scaled spectrum of propylene carbonate at
$158{\rm K}$ \protect\cite{Nagden}. The inset gives a magnified view of the middle section.} \label{fig3}
\end{figure}

\noindent of $\beta $-relaxations. Indeed good fits of the experimental data are possible in this way, including the
lower temperatures, where no shoulder is observed (Fig. 1). In this region it is difficult to determine the $\beta
$-relaxation time unequivocally and therefore it was fixed at values obtained from an extrapolation of the
high-temperature data as explained in detail below. At $102{\rm K}\leq T\leq 106{\rm K}$ deviations of fits and
experimental data show up at $\nu <\nu _{p}$, which can be ascribed to the successive transition of the sample into
the plastic-crystalline state during heating, as mentioned above.

A commonly used description of the excess wing is the so-called Nagel-scaling \cite{Di90a}. For many glass-formers,
the $\varepsilon ^{\prime \prime }(\nu )$-curves for different temperatures and materials can be scaled onto one
master curve by plotting $Y:=1/w\log _{10}[\varepsilon ^{\prime \prime }\nu _{p}/(\Delta \varepsilon \,\nu )]$ {\it
vs.} $X:=1/w(1+1/w)\log _{10}(\nu /\nu _{p})$. Here $w$ denotes the half-width of the loss peak normalized to that of
a Debye-peak \cite{boettcher} and $\Delta \varepsilon $ is the relaxation strength. In Fig. 3 a Nagel-scaling plot of
the present data is shown. In addition, a curve for supercooled propylene carbonate is included, which closely follows
the master curve reported in \cite{Di90a} extending it to higher values of the abscissa \cite {Nagden}. The scaled
data for ethanol show marked deviations from this curve. For $T\geq 126{\rm K}$, the curves are located above the
master curve, due to the contributions from the newly detected $\beta $-relaxation. Similar deviations were observed
in orientationally disordered cyclo-octanol \cite{Molveno} where a $\beta $-relaxation was shown to be responsible for
the apparent excess wing reported in \cite{LP}. In contrast, the scaled curves for $T\leq 102{\rm K}$ fall {\it below}
the master curve, similar to our findings in various orientationally disordered crystals \cite{wingplas}. This
behavior is of special significance as spectra falling above the master curve may always be explained assuming
contributions{\it \ in addition} to the excess wing, but this is not the case for spectra falling below the master
curve. The Nagel-scaling seems to be clearly violated in glass-forming ethanol.

The width parameter $\beta _{CD}$ \cite{boettcher} of the $\alpha $-relaxation, obtained from the fits in Figs. 1 and
2, varies between $0.76$ at $96{\rm K}$ and $1$ at $T>130{\rm K}$. The width parameter $\alpha _{CC}$ \cite{boettcher}
decreases from $0.7$ to $0.4$ with temperature. However, at $T>160{\rm K}$ a clear statement concerning its
temperature development is not possible, as the curves start to shift out of the investigated frequency window. At
$T<110{\rm K,}$ $\beta _{CD}$ and $1-\alpha _{CC}$ are equal to the exponents of the two power laws, $\beta $ and $b$\
observed at $\nu >\nu _{p}$. The predicted relation between theses exponents, $b+1/(\beta +1)\approx 0.72$
\cite{Menon}, is not fulfilled in supercooled ethanol. This could be expected having in mind that this relation was
deduced from the Nagel-scaling, which seems to be violated in glass-forming ethanol (Fig. 3).

\subsection{Relaxation times}

In Fig. 4(a) the relaxation times of the three processes detected in the present work are shown in an Arrhenius
representation. In contrast to the earlier reports \cite{Be98,JR99}, the temperature range investigated in the present
work is more complete, allowing for a detailed analysis of the temperature dependent $\alpha $-relaxation time in
glass-forming ethanol.  The average $\alpha $-relaxation times shown in Fig. 4(a) have partly been determined from the
fits shown  in Figs.

\begin{figure}[tbp]
\begin{center}
\includegraphics[clip,width=8cm]{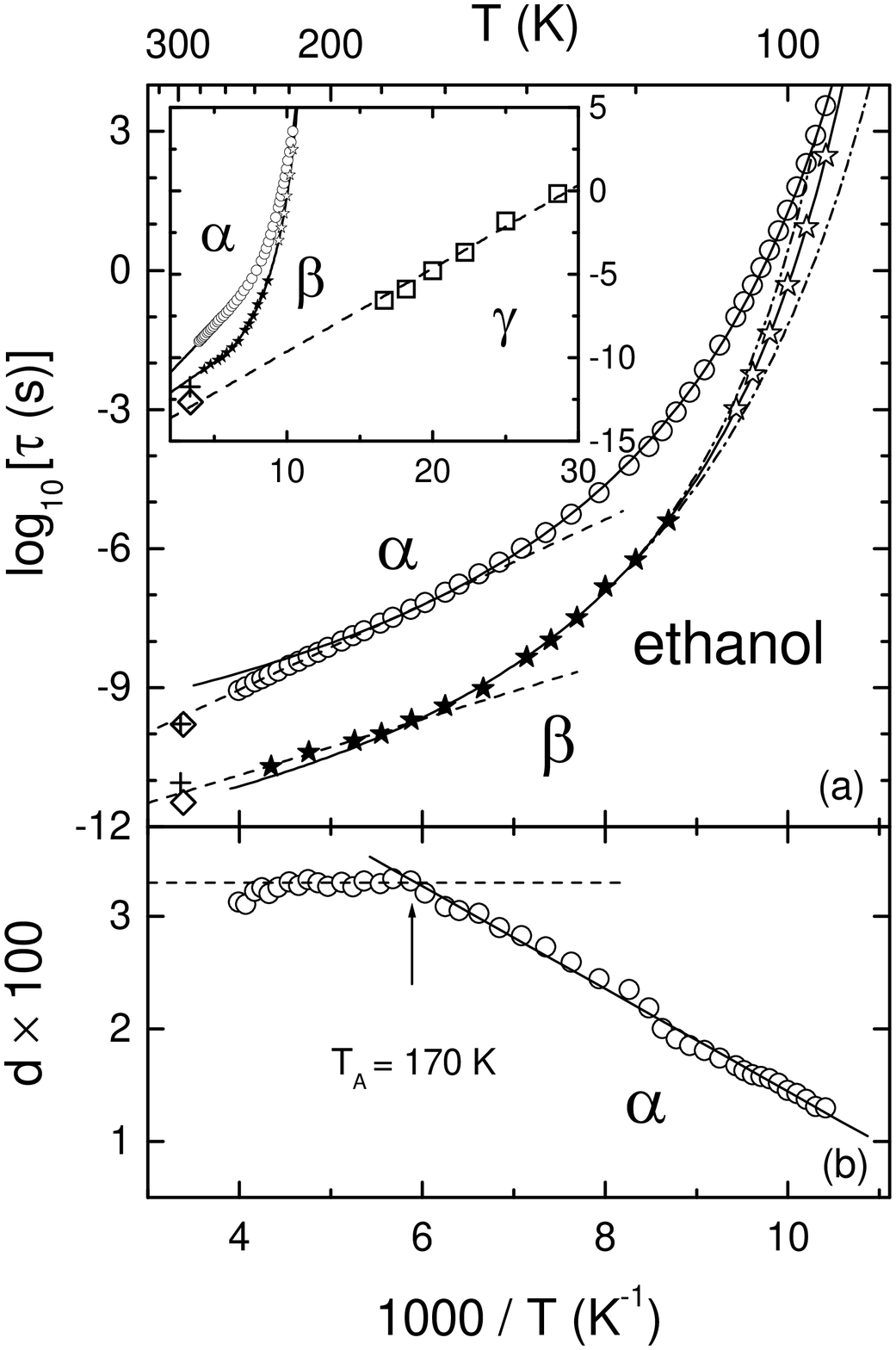}
\end{center}
\caption{(a) Relaxation times of the three processes detected in glass-forming ethanol (the $\protect\gamma $-process
is shown in the inset). The open stars denote $\protect\beta $-relaxation times as obtained from an extrapolation of
the high-temperature fit results. The pluses and lozenges are the results from \protect\cite{Barthel} and
\protect\cite{Kindt}, respectively. The upper solid line shows a VFT fit of $\protect\tau _{\protect\alpha }(T)$ at
$T<T_{A}=170{\rm K}$ and the upper dashed line is an Arrhenius fit at $T>170{\rm K}$. The lower solid line is a VFT
fit of $\protect\tau _{\protect\beta }(T)$ at $T<170{\rm K}$ and the lower dashed line is an Arrhenius fit at
$T>170{\rm K}$. The dash-dotted lines represent lower and upper limits for $\protect\tau _{\protect\beta }(T)$ as
described in the text. The dashed line in the inset is a fit with Arrhenius behavior. For the fit parameters, see
Table 1. (b) Derivative plot of the $\protect\alpha $-relaxation time after Stickel {\it et al.} \protect\cite{Stick1}
(the meaning of d is noted in the text). The solid and dashed lines have been calculated with the same parameters as
in (a). } \label{fig4}
\end{figure}

\noindent 1 and 2 ($\tau _{\alpha }=\tau _{CD}\beta _{CD}$ \cite{boettcher}), performed at selected temperatures only,
and partly calculated from $\tau _{\alpha }\approx 1/(2\pi \nu _{p})$. The latter estimation involves an error of less
than 5\% as long as $\beta _{CD}>0.75$. Often the Vogel-Fulcher-Tammann (VFT) law, $\tau =\tau _{0}\exp
[DT_{VF}/(T-T_{VF})]$ \cite{VFT}, is employed to parameterize the $\alpha $-relaxation time $\tau _{\alpha }(T)$ in
disordered materials, with the Vogel-Fulcher temperature $T_{VF}$\ and the strength parameter $D$ \cite {strength}.
Indeed at temperatures $T<T_{A}=170{\rm K}$, $\tau _{\alpha }(T)$ of glass-forming ethanol [circles in Fig. 4(a)] can
be well described in this way (the parameters of this and the other fits shown in Fig. 4 are collected in Table 1). At
high temperatures, $\tau _{\alpha }(T)$ shows deviations from a VFT law. In Fig. 4(a) they were taken into account by
assuming a transition to thermally activated behavior (dashed line) \cite{Stick2}. This notion finds support in the
derivative plot after Stickel {\it et al.} \cite{Stick1} [Fig. 4(b)], where $d:=[-d(\log _{10}\nu
_{p})/d(1/T)]^{-%
{\frac12}%
}$\ is plotted {\it vs.} the inverse temperature. This plot leads to a linearization of the VFT law and to a constant
for thermally activated behavior. A similar analysis was performed in \cite{Stick2} where $T_{A}=165{\rm K}$ was
deduced in reasonable agreement with the present value. However, one should mention that the use of derivatives for
the test of fitting formulas and theoretical predictions can be criticized \cite{Kiv}. Based on various theoretical
models of the glass-transition, many alternative descriptions of $\tau _{\alpha }(T)$-curves in glass-forming
materials have been proposed, e.g. in \cite{theos}. They often lead to fits of similar quality \cite{PCPRE,Cummins}\
and often it is difficult to arrive at a decision in favor or against a specific model from the analysis of the
$\alpha $-relaxation time. In Fig. 4(a) we also include results from microwave \cite{Barthel} and far-infrared
investigations \cite{Kindt} of liquid ethanol at room-temperature. The spectra in \cite{Barthel,Kindt} were analyzed
assuming a sum of three Debye-relaxations. The slowest relaxation time obtained in this way is in reasonable accord
with an extrapolation of the high-temperature Arrhenius-law used for the description of our data.

In the inset of Fig. 4 the $\gamma $-relaxation time is included. It agrees reasonably with the results reported
earlier \cite{Be98,JR99} and is consistent with the fastest relaxation time reported in \cite{Barthel,Kindt}. The
dashed line is a fit of the present data and the room-temperature point from \cite{Kindt} using an Arrhenius law.

The filled stars in Fig. 4(a) represent the $\beta $-relaxation time $\tau _{\beta }(T)$ obtained from the fits of
$\varepsilon ^{\prime \prime }(\nu )$ with the sum of a CD and a CC function (Figs. 1 and 2). Again the present
results can be reasonably extrapolated to the relaxation times of the second-fastest process reported in
\cite{Barthel,Kindt} by assuming a thermally activated behavior above $170{\rm K}$ (dashed line). Towards lower
temperatures, $\tau _{\beta }(T)$ significantly deviates from an Arrhenius behavior and can be described by a VFT law
(solid line). At low temperatures, where a shoulder is no longer observable (Fig. 1), it is difficult to unequivocally
determine $\tau _{\beta }$ from the fits. Therefore for the fits shown in Fig. 1 at $T\leq 106{\rm K}$, $\tau _{\beta
} $ was fixed at values obtained from an extrapolation of the VFT law towards low temperatures [open stars in Fig.
4(a)]. However, it is possible to determine a lower-limit value of $\tau _{\beta }$ by performing fits with $\tau
_{\beta }$, fixed at successively lower values, until intolerable deviations of experimental data and fit occur. The
lower dash-dotted line in Fig. 4(a) is the lowest possible VFT curve taking into account these lower-limit values of
$\tau _{\beta }$. An upper-limit VFT curve was deduced by assuming a maximum possible value of $\tau _{\beta }=\tau
_{\alpha }$ at $T_{g}$ [upper dash-dotted line in Fig. 4(a)].

Irrespective of these uncertainties at low temperatures, $\tau _{\beta }(T)$ exhibits clear deviations from a
thermally activated behavior. Therefore one may have objections to use the term ''$\beta $-relaxation'' for the
relaxation, causing the excess wing in supercooled ethanol, because $\beta $-relaxations are commonly found to follow
an Arrhenius behavior. However, there is no principle reason that $\beta $-processes always should behave thermally
activated, especially as their microscopic origin is still unclear. Already Johari \cite{Johari} suspected that in
systems without a well resolved $\beta $-process the relaxation times of $\alpha $- and $\beta $-process are closer
together due to a uncommon temperature dependence of $\tau _{\beta }$. In some respects, in ethanol the situation is
similar to that in glycerol or propylene carbonate. In these materials, from aging experiments at $T<T_{g}$, we also
found a $\beta $-relaxation as the probable cause of the excess wing \cite{wingbet}. Here the $\beta $-relaxation time
also deviates from thermally activated behavior \cite{Lunkihab} and the $\beta $-relaxation is difficult to detect due
to the lack of a clear separation of $\tau _{\alpha }$ and $\tau _{\beta }$.

In this context it is of interest that recently for glass-formers with a well pronounced $\beta $-relaxation a
correlation of $\tau _{\beta }$ and the Kohlrausch-exponent $\beta _{KWW}$ (describing the width of the
$\alpha$-peak), both at $T_{g}$, was found \cite{Ngaibeta}: $\log _{10}\tau _{\beta }(T_{g})$ increases nearly
linearly with $\beta _{KWW}(T_{g})$. It was noted that those glass-formers that show no well-resolved $\beta
$-relaxation, e.g. glycerol or propylene carbonate, have relatively large values of $\beta _{KWW}(T_{g})$. As the
mentioned correlation implies that the $\alpha $- and $\beta $-timescales approach each other with increasing $\beta
_{KWW}(T_{g})$, it is easily rationalized that in those materials the $\beta $-relaxation is difficult to observe.
Indeed we find a relatively large $\beta _{KWW}(T_{g})=0.82$ for glass-forming ethanol. Using the linear relationship
of log$_{10}\tau _{\beta }(T_{g})$ and $\beta _{KWW}(T_{g})$ given in \cite{Ngaibeta}, leads to the prediction
log$_{10}\tau _{\beta }(T_{g})=15{\rm s}$, which agrees well with the extrapolated VFT law for the $\beta $-process,
shown in Fig. 4(a). In \cite{Ngaibeta} an explanation of this relationship within the coupling model \cite{CM} was
proposed, which also may be consistent with the observed deviation of $\tau _{\beta }(T)$ from thermally activated
behavior.

On the other hand, it should be mentioned that the relaxation-time plot of glass-forming ethanol [Fig. 4(a)] looks
quite similar to that determined for 1-propanol \cite{Hansen,Kudlik}. In this material (and other primary alcohols
\cite{Garg}) also three relaxation processes have been detected by dielectric spectroscopy \cite{Hansen,Kudlik}.
Similar to the present results, the relaxation time of the second process in 1-propanol was found to exhibit marked
deviations from thermally activated behavior. In \cite{Hansen} the explanation for the observed relaxation behavior is
quite different to the picture developed above: The second process was interpreted as the ''true'' $\alpha
$-relaxation. This picture is based on the finding that the relaxation dynamics of the second relaxation in 1-propanol
is paralleled by data obtained from methods coupling to the structural relaxation. The low-temperature/high-frequency
process (denoted as $\gamma $-relaxation in the present work) was assumed to be a Johari-Goldstein $\beta
$-relaxation. In \cite{Hansen} the dominating Debye-type low-frequency process was termed ''$\alpha ^{\prime
}$-relaxation'' and assigned to distinct OH-group motions, but also other explanations were proposed \cite
{Hassion,Floriano}. Interestingly the present $\tau _{\beta }(T)$-data of ethanol are of similar magnitude as (but not
identical to) the average molecular rotation times{\em \ }determined from NMR measurements \cite{eguchi}. However,
they clearly deviate from the results of mechanical spectroscopy \cite{Emery}. Also it is noteworthy that the slowest
process in ethanol exhibits deviations from Debye behavior at low temperatures, in contrast to the ''$\alpha ^{\prime
}$-relaxation'' in 1-propanol.
\begin{table}
\begin{center}
\begin{minipage}{8cm}
\begin{tabular}[t]{c|cc|ccc}
& \multicolumn{2}{|c|}{Arrhenius} & \multicolumn{3}{|c}{VFT} \\ \hline\hline process & $\tau _{0}({\rm s})$ &
$E/k_{B}({\rm K})$ & $\tau _{0}({\rm s})$ & $D$ & $T_{VF}({\rm K})$ \\ \hline
$\alpha $ & $1.9\times 10^{-13}$ & $2110$ & $5.3\times 10^{-11}$ & $8.4$ & $%
76$ \\
$\beta $ & $5.0\times 10^{-13}$ & $1390$ & $2.0\times 10^{-13}$ & $7.9$ & $%
78 $ \\ $\gamma $ & $2.4\times 10^{-15}$ & $1150$ &  &  &
\end{tabular}
\end{minipage}
\end{center}
\caption{Parameters of the fits of the temperature-dependent relaxation times shown in Fig. 4(a).}
\end{table}

\section{SUMMARY}

In the present dielectric investigation of glass-forming ethanol we have found clear evidence for a third, so-far
undetected relaxation process. For the physical origin of the three processes two different scenarios are possible:
The similarity of the relaxation map of ethanol with that found in glass-forming 1-propanol suggests a similar
explanation of these processes as promoted in \cite{Hansen}, especially concerning the identification of the second
process with the structural $\alpha $-relaxation. Alternatively, the second process may be simply a Johari-Goldstein
$\beta $-process with an uncommon non-Arrhenius temperature dependence, similar to our findings in glycerol and
propylene carbonate \cite{wingbet}. However, the main result of the present study remains unaffected by these open
questions: The apparent excess wing in glass-forming ethanol \cite{Be98,JR99}, is due to the newly detected relaxation
process. This finding further corroborates the notion that the excess wing is not a separate feature in the spectra of
glass-formers, but can be commonly ascribed to additional relaxation processes \cite{wingbet}. In addition, we can
state a clear violation of the Nagel-scaling \cite{Di90a} in glass-forming ethanol, at least if the parameters of the
slowest relaxation are used for the scaling. Of course this may be unjustified, if the scenario analogous to
1-propanol \cite{Hansen} is correct. In this case it is difficult to make a statement about the Nagel-scaling due to
the interference of the ''$\alpha $-relaxation'' with the ''$\alpha ^{\prime }$-relaxation'' at low and the
''$\beta$-relaxation'' at high frequencies. Finally, in the light of our finding of the absence of the excess wing in
orientationally disordered crystals \cite{wingplas}, it certainly would be of interest to check also for the presence
of a third process in the plastic crystalline phase of ethanol. Such measurements are currently in progress and will
be reported in a future publication.

\acknowledgments

We thank S. Benkhof for calling our attention to dielectric measurements in ethanol. This work was financially
supported by the Deutsche Forschungsgemeinschaft, Grant-Nos. LO264/8-1 and LO264/9-2 and partly by the BMBF,
contract-No. EKM 13N6917.

\end{document}